\documentclass[prd,nofootinbib,floatfix,superscriptaddress,twocolumn,showpacs,reprint]{revtex4-1}
\usepackage{graphicx,wrapfig,float,slashed,subcaption}
\usepackage{amsmath,amssymb,epsfig,graphicx}
\usepackage{multirow}
\usepackage{ragged2e}
\usepackage{array}

\newcommand{\nc}{\newcommand}
\nc{\non}{\nonumber}
\nc{\hc}{\hbox {H.c.}}
\nc{\noi}{\noinde	nt}
\nc{\barx}{\bar{x}}
\nc{\pbarn}{\;\hbox {pb}}
\nc{\fbarn}{\;\hbox {fb}}
\nc{\ra}{\rightarrow}
\nc{\met}{p_{T}^{\textnormal{miss}}}
\nc{\lep}{\ell}
\nc{\gev}{\textnormal{GeV}}
\usepackage{color}
\usepackage[usenames,dvipsnames,table]{xcolor}
\usepackage[colorlinks=true
,urlcolor=Maroon
,anchorcolor=Blue
,citecolor=OliveGreen
,filecolor=Blue
,linkcolor=NavyBlue
,menucolor=Blue
,pagecolor=Blue
,linktocpage=true
,pdfproducer=medialab
,pdfa=true
]{hyperref}

\usepackage{cleveref}

\definecolor{agray}{rgb}{0.95, 0.95, 0.99}

\nc{\hsp}{\hspace{0.5cm}}
\nc{\lsp}{\hspace{1cm}}
\nc{\Lsp}{\hspace{2cm}}
\nc{\LLsp}{\lsp\lsp}
\nc{\lra}{\longrightarrow}
\nc{\p}{\prime}
\nc{\sgn}{\text{sgn}}
\nc{\ph}{\varphi}
\nc{\beq}{\begin{equation}}  \nc{\eeq}{\end{equation}}
\nc{\bea}{\begin{eqnarray}}  \nc{\eea}{\end{eqnarray}}
\nc{\baa}{\begin{array}}     \nc{\eaa}{\end{array}}
\nc{\bit}{\begin{itemize}}   \nc{\eit}{\end{itemize}}
\nc{\ben}{\begin{enumerate}} \nc{\een}{\end{enumerate}}
\nc{\bce}{\begin{center}}    \nc{\ece}{\end{center}}
\nc{\bpm}{\begin{pmatrix}}   \nc{\epm}{\end{pmatrix}}
\nc{\bvt}{\begin{verbatim}}  \nc{\evt}{\end{verbatim}}

\def\lsim{\mathrel{\raise.3ex\hbox{$<$\kern-.75em\lower1ex\hbox{$\sim$}}}}
\def\gsim{\mathrel{\raise.3ex\hbox{$>$\kern-.75em\lower1ex\hbox{$\sim$}}}}

\def\udots{\mathinner{\mkern1mu\raise1pt\vbox{\kern7pt\hbox{.}}\mkern2mu\raise4pt\hbox{.}\mkern2mu\raise7pt\hbox{.}\mkern1mu}}

\def\gev{\;\hbox{GeV}}

\def\fb{\;\hbox{fb}}

\begin{document}

\title{Observability of Inert Scalars at the LHC}

\author{Majid Hashemi} 
\email{hashemi\_mj@shirazu.ac.ir}
\affiliation{Physics Department, College of Sciences,
Shiraz University, Shiraz, 71946-84795, Iran}

\author{and Saereh Najjari}
\email{saereh.najjari@fuw.edu.pl}
\affiliation{Faculty of Physics,
University of Warsaw,
Pasteura 5, 02-093 Warsaw, Poland}

\date{\today}

\begin{abstract}
In this work we investigate the observability of Inert Doublet Model scalars at the LHC operating at the center of mass energy of 14  TeV. The signal production process is $pp \to AH^\pm \to ZHW^{\pm}H$ leading to two different final states of $\ell^+\ell^-Hjj H$ and  $\ell^+\ell^-H\ell^\pm\nu H$ based on the hadronic and leptonic decay channels of the W boson. All the relevant background processes are considered and an event selection is designed to distinguish the signal from the large Standard Model background. We found that signals of the selected search channels are well observable at the LHC with integrated luminosity of 300$\fb^{-1}$. 
\end{abstract}

\maketitle
\flushbottom


\section{Introduction}
\label{Introduction}
Inert Doublet Model (IDM) is a special type of  the  Two Higgs Doublet Model (2HDM) that respects a discrete $Z_2$ symmetry under
which the Standard Model (SM) fields are {\it even} while the inert doublet additional $SU(2)$  scalar doublet $\Phi_D$ is {\it odd}, therefore the neutral component of $\Phi_D$ could be a dark matter candidate \cite{Deshpande:1977rw,Cao:2007rm,Barbieri:2006dq,LopezHonorez:2006gr,Honorez:2010re,Dolle:2009fn,Goudelis:2013uca}.  Only the SM Higgs doublet acquires a non-zero vacuum
expectation value and hence is a source of electroweak symmetry
breaking (EWSB). The inert doublet does not couple with the fermions of the
SM. After EWSB in the scalar sector this model has five
physical states: the SM-like Higgs boson $h$ as well as two
charged scalars, $H^\pm$, and two neutral ones, $H$ and $A$.

The inert doublet scalars have been studied in \cite{Aoki:2013lhm} aiming at their mass reconstruction at a linear collider. A detailed study recently presented in \cite{Hashemi:2015swh} shows that all inert scalars can well be identified and their masses can be measured up to a reasonable accuracy at a future linear collider. The observation of Inert scalars at the LHC has also attracted attention. For a list of recent works one may refer to \cite{Ilnicka:2015jba, Blinov:2015qva, Arhrib:2013ela, Belanger:2015kga, Poulose:2016lvz, Datta:2016nfz}. These studies have been done mostly based on the production of $H^+H^-$, $HA$, $HH^+$ and $AH^+$ followed by the $H^+ \to W^+ H$ and $A\to Z H$ assuming $H$ as the dark matter candidate. In one of the most recent works in the list (\cite{Poulose:2016lvz}) a di-jet plus missing transverse energy signature of the inert doublet model events has been studied. The studied benchmark points (which are different from ours) are shown to be observable at a minimum integrated luminosity of 500  fb$^{-1}$. There has been also a study of IDM trilepton signals at the LHC in \cite{Miao:2010rg} at an integrated luminosity of 300  fb$^{-1}$. However, the single electroweak gauge boson production ($W$+jets and Drell-Yan $Z/\gamma$+jets) has not been considered.
 
In this work, a scenario is considered in which the scalar $H$ is the dark matter candidate, i.e., $m_H <m_{H^\pm}, m_A$.  We propose a new set of  benchmark points which satisfy all the recent experimental and theoretical constraints. The analysis is performed in two categories: {\it di-lepton plus di-jet} and {\it tri-lepton} final states, all of them produced through $A^0 H^\pm$  at the center of mass energy of 14 TeV. The lepton denoted by $\ell$ is either an electron or a muon. Therefore signal events include both electrons and muons in the final states.  
The following processes are considered as the signal in our studies:
\begin{align}
q \bar{q} &\to W^\pm\to AH^\pm \to Z^{(*)} H W^{\pm(*)}H \to \lep^+ \lep^-H jj (\ell^\pm \nu)H.
\label{yukawa_lagrangian_2}
\end{align}%
Concerning the decay channels, $H^\pm$ decays to $W^\pm H$ and $A$ decays to $ZH$, however, at all the benchmark points under study, $M_{H^\pm}< (M_{W^\pm}+M_H)$ and $M_A< (M_Z+M_H)$,  therefore those decays can only occur via a virtual $W^{\pm*}$ and  $Z^*$ boson. 
 After introducing  in \cref{Inert Doublet Model} the model  and setting up the notations , in  \cref{Benchmark Points} we provide the benchmark points for this study. The simulation tools  used for the analysis described in  \cref{Software Setup}. The event generation and the analysis of our benchmark points are given in \cref{Signal and Background processes,Signal and Background Cross sections,Event Selection and Analysis}. Finally, we  discuss and conclude in \cref{Discussion,Conclusions}.


\section{Inert Doublet Model}
\label{Inert Doublet Model}
IDM  is an extension of scalar sector of the SM by addition of a scalar doublet $\Phi_D$ to the SM-like Higgs doublet ($\Phi_S$). The inert doublet is odd under the discrete $Z_2$ symmetry, whereas all of the SM fields are even.
The two scalar doublets can be written as,
\beq
\Phi_S=\frac1{\sqrt2}\bpm \sqrt2G^\pm\\ 
v+h+iG^0 \epm, \hspace{0.5cm} \Phi_D=\frac1{\sqrt2}\bpm \sqrt2H^\pm\\H+iA \epm,	\label{higgs_vev}
\eeq
where  $v=246$~GeV denotes the vacuum expectation value of the SM-like Higgs doublet.
The scalar potential for the IDM reads:
\begin{align}
&V(\Phi_S,\Phi_D)=-\frac{1}{2}\Big[m_{11}^2(\Phi_S^\dagger\Phi_S)+m_{22}^2(\Phi_D^\dagger\Phi_D)\Big]\notag\\
&+\frac{\lambda_1}{2}(\Phi_S^\dagger\Phi_S)^2+\frac{\lambda_2}{2}(\Phi_D^\dagger\Phi_D)^2 +\lambda_3(\Phi_S^\dagger\Phi_S)(\Phi_D^\dagger\Phi_D)\notag\\
&+\lambda_4(\Phi_S^\dagger\Phi_D)(\Phi_D^\dagger\Phi_S)
+\frac{\lambda_5}{2}\Big[(\Phi_S^\dagger\Phi_D)^2+(\Phi_D^\dagger\Phi_S)^2\Big].	\label{potential}
\end{align}
The masses and interactions of scalar section are fixed by parameters, ($m_{11,22,},\lambda_{1,2,3,4,5}$). After EWSB  the  physical masses of scalars  are expressed as:
\begin{align}
m_h^2&=\lambda_1 v^2=m_{11}^2, 	\notag\\
m_{H^+}^2&=\frac{1}{2}(\lambda_3 v^2-m_{22}^2),		\notag\\
m_H^2&=\lambda_{L}v^2-\frac{1}{2}m_{22}^2, 	\notag\\
m_A^2&=\lambda_{S}v^2-\frac{1}{2}m_{22}^2,	\label{4d_cc}
\end{align}
with $\lambda_{L,S}$ defined as, $\lambda_{L,S}\equiv \frac{1}{2}(\lambda_3+\lambda_4\pm\lambda_5)$.
The scalar and pseudoscalar mass splitting is related to $\lambda_5$:
\beq
m_H^2-m_A^2=\lambda_{5}v^2
\eeq
Note that in order to have the neutral scalar $H$ to be the lightest scalar of dark sector one requires $\lambda_5< 0$, which will be the case in our analysis.

The theoretical and experimental constraints on IDM reduce the parameter space considerably. In the following two subsections we outline these constraints and then in the next section we propose our benchmarks points which respect all the following constraints.
\subsection{Theoretical Constraints}

We enlist below the theoretical constraints on the IDM parameters:
\begin{itemize}
\item[({\it i})] {\it Stability}: The tree level vacuum stability constraints the potential quartic coupling parameters as:
\beq \lambda_1\geq 0,~\lambda_2\geq 0,~\sqrt{\lambda_1\lambda_2}+\lambda_3>0,~~
\sqrt{\lambda_1\lambda_2}+2\lambda_{L}>0.
\eeq
\item[({\it ii})] {\it global minimum}:  In order to have a global minimum of the inert potential, we impose the following bound on the mass parameters \cite{Swiezewska:2012ej}:
\beq 
\frac{m_{11}^2}{\sqrt{\lambda_1}}\geq \frac{m_{22}^2}{\sqrt{\lambda_2}}.
\eeq
\item[({\it iii})] {\it Tree-level unitarity}: The  constraints from the perturbative unitarity of $2 \to 2$ scattering of the vector bosons in the SM is taken into account. 
\item[({\it iv})] {\it Perturbativity}: We also require that all couplings remain perturbative, i.e. we take 4$\pi$ as upper limits. 
\end{itemize}

\subsection{Experimental Constraints}
\begin{itemize}
\item[({\it i})] {\it SM-like Higgs}:  We employ the SM-like Higgs boson mass $h$ to be $M_h=125$~GeV \cite{ Aad:2015zhl}
and  its total width to be $\Gamma_{h}\leq 22$~MeV \cite{CMS, ATLAS}.
\item[({\it ii})] {\it Gauge boson width bound}: The width of the SM gauge bosons W and  Z   put the following constraint on the mass parameter  \cite{Agashe:2014kda}:
\begin{align}
&m_H+m_A\geq m_Z, &2m_{H^\pm}\geq m_Z ,\notag\\
&m_A+m_{H^\pm}\geq m_W, &m_H+m_{H^\pm}\geq m_W. \label{posi}
\end{align}
\item[({\it iii})] {\it Dark matter}: We take into account the following DM searches bounds:
\begin{itemize}
\item We consider 2$\sigma$ LUX2016 experimental direct searches exclusion bound on the dark matter scattering cross section~\cite{Akerib:2016vxi}.
\item We also employ the Planck measurement  2$\sigma$ limits on dark matter relic density, $\Omega_c h^2=0.1197\pm0.0022$ \cite{Ade:2015xua}.
\end{itemize}
\item[({\it iv})] {\it Charged scalar}:
\begin{itemize}
\item LEP limit on the charged scalar mass of $m_{H^\pm}\geq 70$~GeV \cite{Pierce:2007ut} is taken into account. Moreover, the
exclusion bounds from SUSY searches at LHC and LEP \cite{Lundstrom:2008ai,Belanger:2015kga} are also considered.
\item Limit on the charged scalar total width, $\Gamma_{tot}\geq 6.58 \times 10^{-18}$~GeV, to avoid bounds from long lived charged particle searches \cite{Ilnicka:2015jba}.
\end{itemize}
\item[({\it v})] {\it EWPO}: Electroweak precision observables (EWPO) bound at 2$\sigma$ level is also considered
~\cite{Altarelli:1990zd,Peskin:1990zt,Maksymyk:1993zm,Peskin:1991sw}.
\end{itemize}

\section{Benchmark Points}
\label{Benchmark Points}
\begin{figure*}[t]
\centering
\begin{tabular}{cc}
  \includegraphics[width=0.45\textwidth]{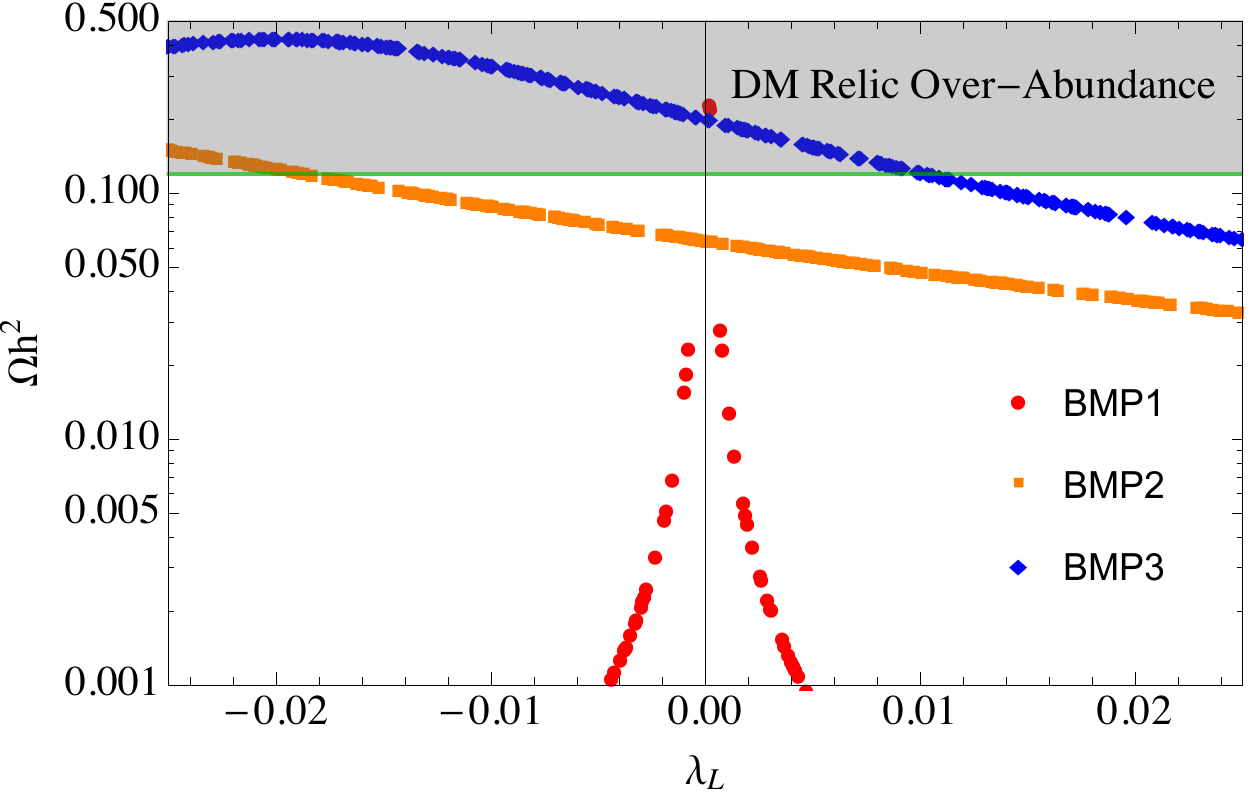}
\hspace{0.5cm}&\hspace{0.5cm}
 \includegraphics[width=0.47\textwidth]{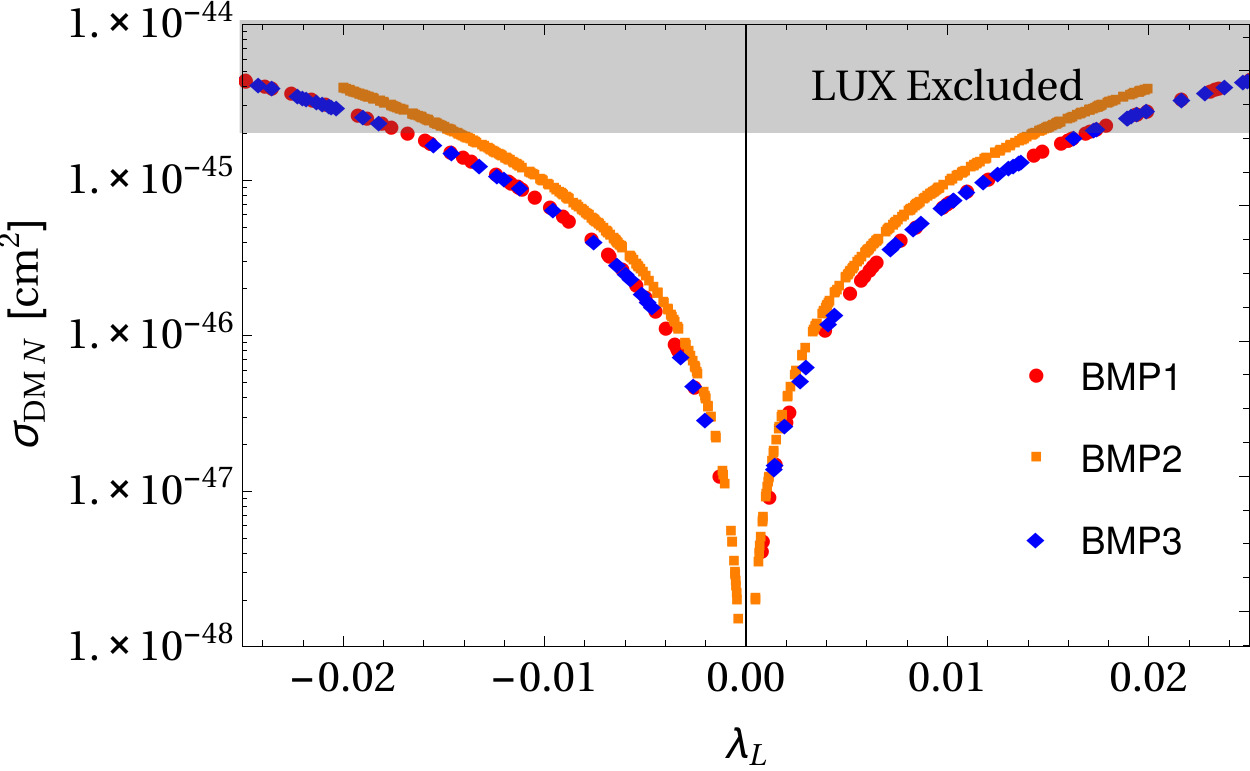}\\
(a)&\hspace{1cm}(b)
\end{tabular}
\caption{(a) Abundance of IDM dark matter as a function of parameter $\lambda_L$. (b) Limits on the dark matter-nucleon cross section as a function of $\lambda_L$.}
\label{constraints}
\end{figure*}

The IDM has seven free parameters, out of which two ($m_{11}$ and $\lambda_1$) are fixed to get the electroweak vev $v=246$~GeV and the SM-like Higgs mass $m_h=125$~GeV. The remaining five parameter $m_{22}, \lambda_{2,3,4,5}$ can be extracted for the physical parameters namely, $m_H,m_A, m_{H^\pm}, \lambda_2$ and $\lambda_L$.
Taking into account all the above mentioned theoretical and experimental bounds, we propose the following benchmark points (BP) which have the potential for observation at the LHC. 
\begin{center}
\begin{tabular}{llll}
 BP1:  \it  &$m_H$= 60 GeV,& $m_A$=111 GeV, &$m_{H^\pm}$=123 GeV,\\
 BP2: \it  &$m_H$= 75 GeV,& $m_A$=120 GeV, &$m_{H^\pm}$=115 GeV,\\
 BP3 : \it &$m_H$= 70 GeV,& $m_A$=110 GeV,& $m_{H^\pm}$=130 GeV.\\
\end{tabular}
\end{center}
Note that the above BPs only fix the inert scalar masses, whereas the two parameters $\lambda_2$ and $\lambda_L$ are free. However, as noted above, there are theoretical and experimental bounds constraining the parameters $\lambda_{2}$ and $\lambda_{L}$. Considering the perturbativity and the tree-level unitarity constraints $\lambda_{2}\leq 4.2$.  
Moreover, $\lambda_{L}<-0.5$ is excluded from the bounds on positivity. There is also a lower bound on $\lambda_{2}$ from positivity \eqref{posi}
We fix the parameter $\lambda_{L}$ in a way that the dark matter candidate $H$ provide the correct dark matter relic density in the universe, i.e. $\Omega_c h^2\simeq 0.1241$. Note that the the parameter $\lambda_{2}$, quadratic self coupling of the inert scalars, does not affect on the value of relic density because mostly the annihilation to the SM is mediated through the SM-like Higgs boson and the parameter $\lambda_L$.  We compute the relic density of dark matter (H) as a function of $\lambda_L$, our result for the benchmark points are summarized in Fig.~\ref{constraints}(a). The direct detection exclusion from the LUX experiment is shown as a function of $\lambda_{L}$ in Fig.~\ref{constraints}(b). We note that $\lambda_{L}\simeq 0.001$ for BP1, $\lambda_{L}\simeq -0.01$ for BP2 and $\lambda_{L}\simeq 0.01$ for BP3 fixes the right dark matter relic abundance and is not excluded by the LUX experiment, $\lambda_{2}=[0,\frac{4\pi}{3}]$ are in agreement with all above mentioned constraints. We used {\tt micrOMEGAs 4.1} \cite{Belanger:2014vza} for the LUX bound and relic density.

\section{Software Setup}
\label{Software Setup}
The generation of signal events starts with implementing the inert doublet model Lagrangian in {\tt LanHEP-3.2.0} \cite{lanhep1,lanhep2}. The {\tt LanHEP} output (model files) are used in {\tt CompHEP-4.5.2} \cite{comphep1,comphep2} for event generation. The hard processes generated by {\tt CompHEP} are then passed to {\tt PYTHIA-8.2.15} \cite{Sjostrand:2014zea} for final state showering and multi particle interactions. The background events are all generated by {\tt PYTHIA}. The analysis is carried out using {\tt ROOT-5.34.30}\cite{Brun:1997pa}.

\section{Signal and Background processes}
\label{Signal and Background processes}
The signal processes are categorized into two main channels. These channels have been selected carefully with the aim of a reasonable background suppression.  The di-lepton plus di-jet channel benefits from a tool which we call ``$W/Z$ veto'' and is powerful against Drell-Yan and any other SM background which has a genuin $W$ boson in the event. The trilepton channel benefits from ``jet veto'' and three lepton requirement which suppresses the Drell-Yan background and any other SM background with jets involved, except for the irreducible $WZ$ process. In what follows, the above two channels are described in detail.   
\subsection{$AH^\pm \to \ell^+\ell^-H jj H$}
\begin{figure*}[t]
\centering
\begin{tabular}{cc}
  \includegraphics[width=0.3\textwidth]{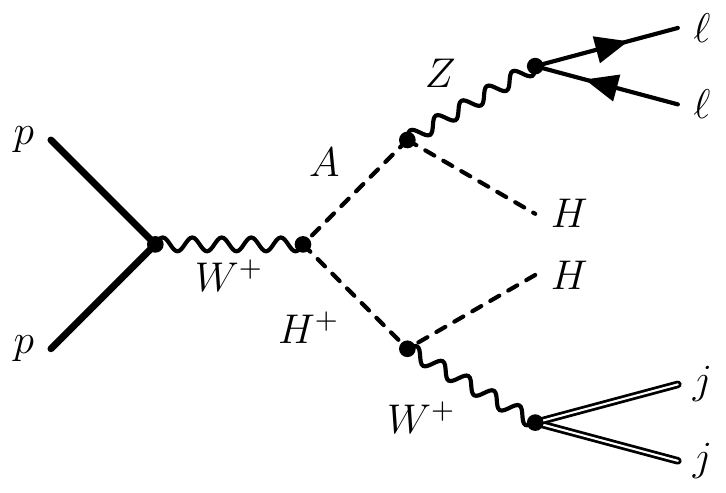}
  \hspace{1cm}&\hspace{1cm}
  \includegraphics[width=0.3\textwidth]{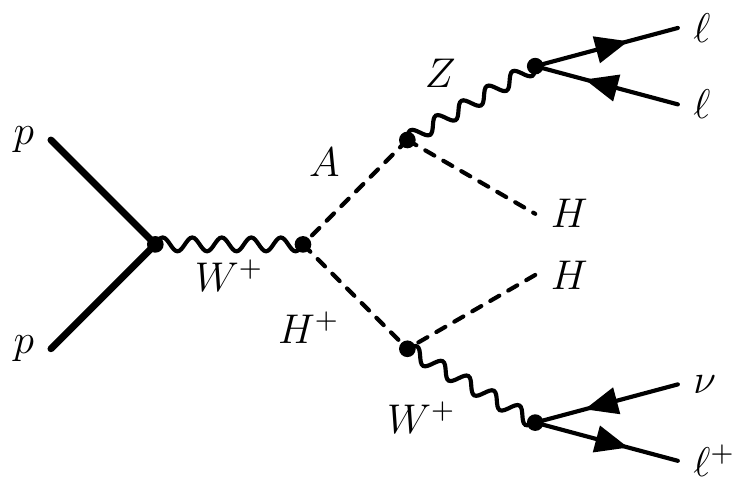}
\\
(a)&(b)
\end{tabular}
\caption{The signal production chain in the di-lepton plus di-jet channel (a), and the trilepton channel (b).}
\label{feyn}
\end{figure*}

The di-lepton plus di-jet channel proceeds through $ AH^\pm$ production followed by the $A\to ZH$ and $H^\pm\to W^\pm H$ decays. The $H$ escapes the detector and can only be seen as missing (transverse) energy. The off-shell bosons decay through $W^\pm \to jj$ and $Z \to \ell \ell$ thus producing a final state consisting of two jets and two leptons. Figure~\ref{feyn}(a) illustrates the signal production chain. The main difference between the signal and the SM background events is the virtuality of gauge bosons which results in low values of di-lepton and di-jet invariant mass distributions far from SM peaks corresponding to the $Z$ and $W$ bosons at roughly 90 and 80 GeV. A reasonable background suppression is achieved by requiring missing transverse energy threshold and exactly two leptons and two jets in the event. The two jets should be light ($u, d, s$) for $t\bar{t}$ suppression and should not have an invariant mass near the  $W$ boson mass to reject the SM gauge boson pair or single production. The $W/Z$ boson ``veto'' suppresses both $W$+jets and Drell-Yan $Z^{(*)}/\gamma^*$+jets dramatically.

\subsection{$AH^\pm \to \ell\ell H \ell \nu H$}  
The trilepton channel is produced through $AH^\pm$ process with $A\to ZH$ and $H^\pm\to W^\pm H$ decays. Both $Z$ and $W$ bosons decay leptonically in this case. Therefore the final state consists of three leptons, two of which are expected to give an invariant mass near the off-shell $Z^*$ boson mass in the signal. Figure~\ref{feyn}(b) shows the signal production chain. Since there is no jets in the signal process, SM background events, such as $t\bar{t}, ~W+\textnormal{jets},~Z/\gamma+\textnormal{jets}$ are well suppressed by the jet veto and requiring exactly three leptons in the event. Since the final aim in this channel is the distribution of di-lepton from the $Z^*$ boson, the Drell-Yan background should be suppressed down to a reasonable level. It is, in fact, dramatically suppressed by requiring three leptons in the event. The main background will then be the irreducible $WZ$ production which produces the same type of final state particles. This background is also shown to be well under control by requiring hard leptons above a reasonable threshold ($E_{T}>30$ GeV).   

\section{Signal and Background Cross sections}
\label{Signal and Background Cross sections}
The signal cross section is calculated using {\tt CompHEP} at event generation time. Tabs. \ref{sxsec} and \ref{sbr} show cross sections of the signal process ($AH^{\pm}$) as well as branching ratios of relevant decays for three benchmark points. The cross sections of background processes are also presented in Tab. \ref{bxsec}.
\begin{table}[h]
\centering
\begin{tabular}{|c|c|c|c|}
\hline
\multirow{2}{*}{Process} & \multicolumn{3}{c|}{$H^\pm A$}  \\
& BP1 & BP2 & BP3     \\
\hline\hline
Cross section [fb] & 255 & 244 & 232     \\
\hline
\end{tabular}
\caption{Signal ($AH^\pm$) cross sections at $\sqrt{s}=14$ TeV. \label{sxsec}}
\end{table}
\begin{table}[h]
\centering
\begin{tabular}{||c|c|c|c|c||}
\hline
Process&$H^+\to W^+ H$  &$H^+\to W^+ A$ &$A\to Z H$ & $A\to W^+ H^-$\\
\hline\hline
BP1 &$0.99$ &$1.44\times 10^{-4}$  &1&0  \\
\hline
BP2 &1 & 0 &$0.99$& $3.94\times 10^{-5}$ \\
\hline
BP3 &$0.99$ &$5.05\times 10^{-3}$  &1& 0 \\
\hline
\end{tabular}
\caption{Branching ratio of charged and neutral scalar decays. \label{sbr}}
\end{table}

\begin{table}[h]
\centering
\begin{tabular}{|c|c|c|c|c|c|c|c|c|}
\hline
Process & $t\bar{t}$ & WW & WZ & ZZ & STS & STT & W & Z \\
\hline
Cross section [pb] & 800 & 70.6 & 25.7 & 11.2 & 8.27 & 172.4 & 1.5e5 & 8.3e4 \\
\hline
\end{tabular}
\caption{Background cross sections at $\sqrt{s}=14$ TeV. ``STS'' and ``STT'' denote the single top $s$ and $t$ channels respectively. \label{bxsec}}
\end{table}

\section{Event Selection and Analysis}
\label{Event Selection and Analysis}
In this section, event selection strategy and details of the analysis are presented. Before proceeding to the details, it should be mentioned that all leptons and jets four momenta are smeared according to LHC results reported in \cite{smear1,smear2}. The smearing is based on a gaussian distribution with a width of 15$\%$ of the jet energy and 2$\%$ of the lepton transverse momentum. These values are equivalently the jet (lepton) energy (transverse momentum) resolutions.  
\subsection{The dilepton plus di-jet channel}
This signal consists of two leptons and two jets. Eq. \ref{eeta2l2j} shows the kinematic cut applied on all leptons and jets in the event. Therefore all leptons and jets are required to pass a threshold of 30 GeV applied on their transverse momenta and they are required to be in the central barrel and endcap regions and not outside the $|\eta|<3$ region. Here $\eta$ is defined as $\eta=-\ln\tan(\theta/2)$ with $\theta$ being the polar angle.
\beq
E_{T}^{\textnormal{jet/lepton}}> 30 ~\textnormal{GeV},~~|\eta|<3.
\label{eeta2l2j}
\eeq
This requirement is basically useful for SM background suppression when an event involves soft leptons or jets like in a single or pair production of gauge bosons. If a lepton or jet passes this requirement, it is counted. 

The $b$-tagging is also performed based on a simple spatial matching between the $b$-jet and the true $b$ or $c$ quark in the event. If a $b$-jet lies in the vicinity of a $b$ or $c$ quark, i.e., $\Delta R (b\textnormal{-jet}, b/c~\textnormal{quark})<0.4$, it is accepted by 60$\%$ or 10$\%$ probability respectively. If a jet is not identified as a $b$-jet, it is taken as a light jet. An event is required to have exactly two light jets and two leptons passing the requirement in Eq. \ref{eeta2l2j} to be selected for further analysis. 

\begin{figure}[h]
\centering
  \includegraphics[width=0.5\textwidth]{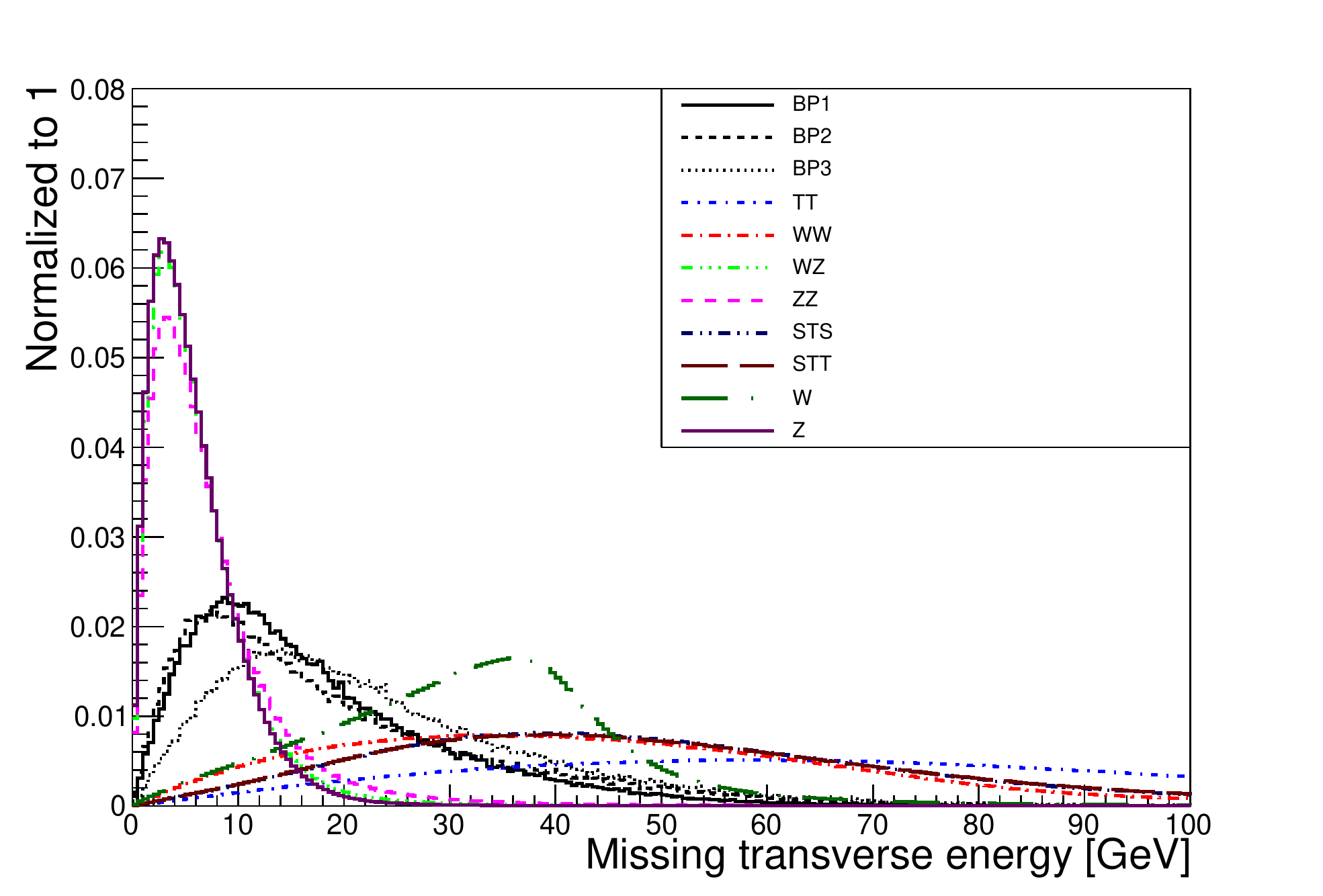}
  \caption{The distribution of missing transverse energy in signal ($\ell\ell H jj H$) and background events.}
  \label{met2l2j}
\end{figure}

\begin{figure*}[t]
\centering
\begin{tabular}{cc}
  \includegraphics[width=0.47\textwidth]{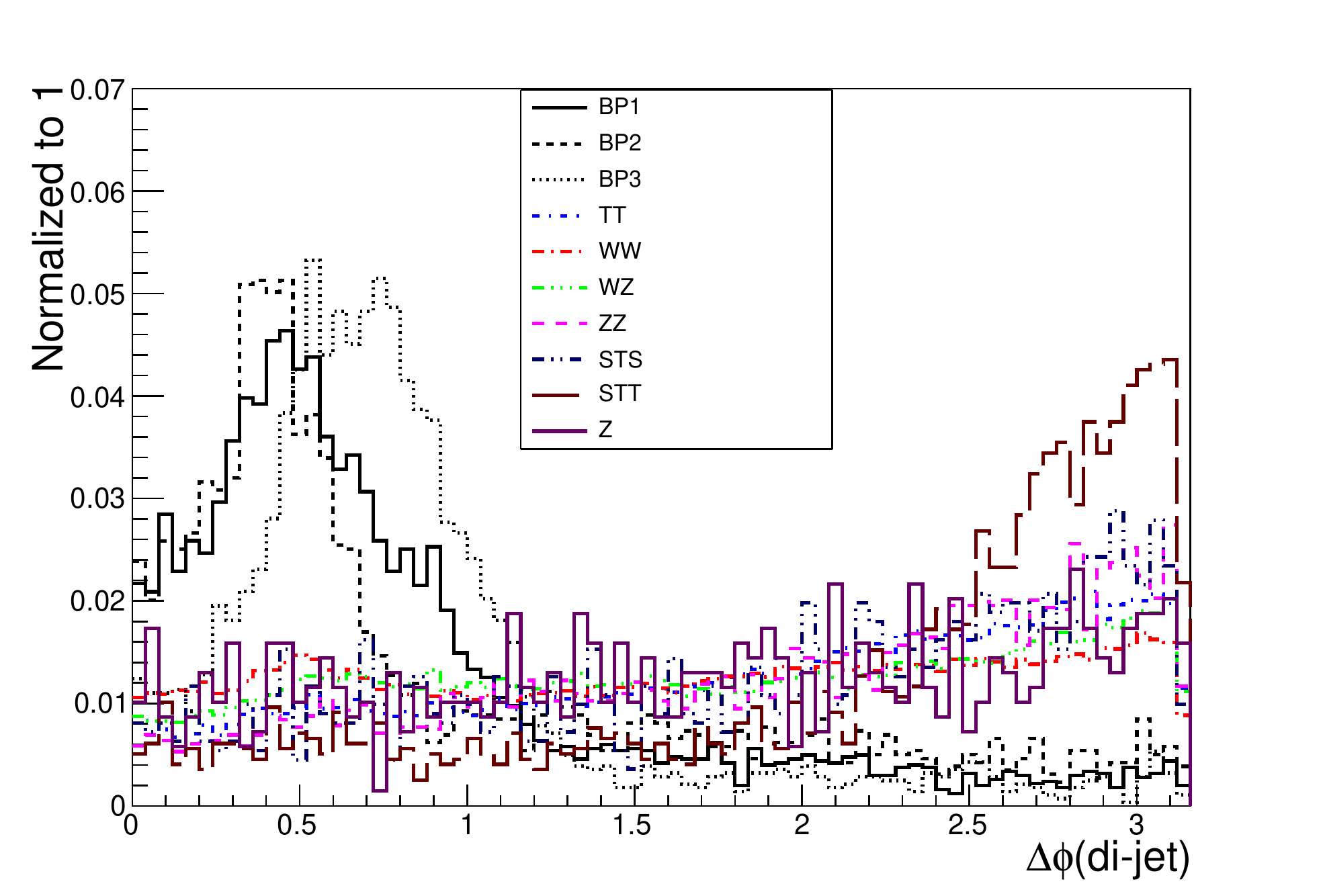}
\hspace{0.5cm}&\hspace{0.5cm} 
 \includegraphics[width=0.47\textwidth]{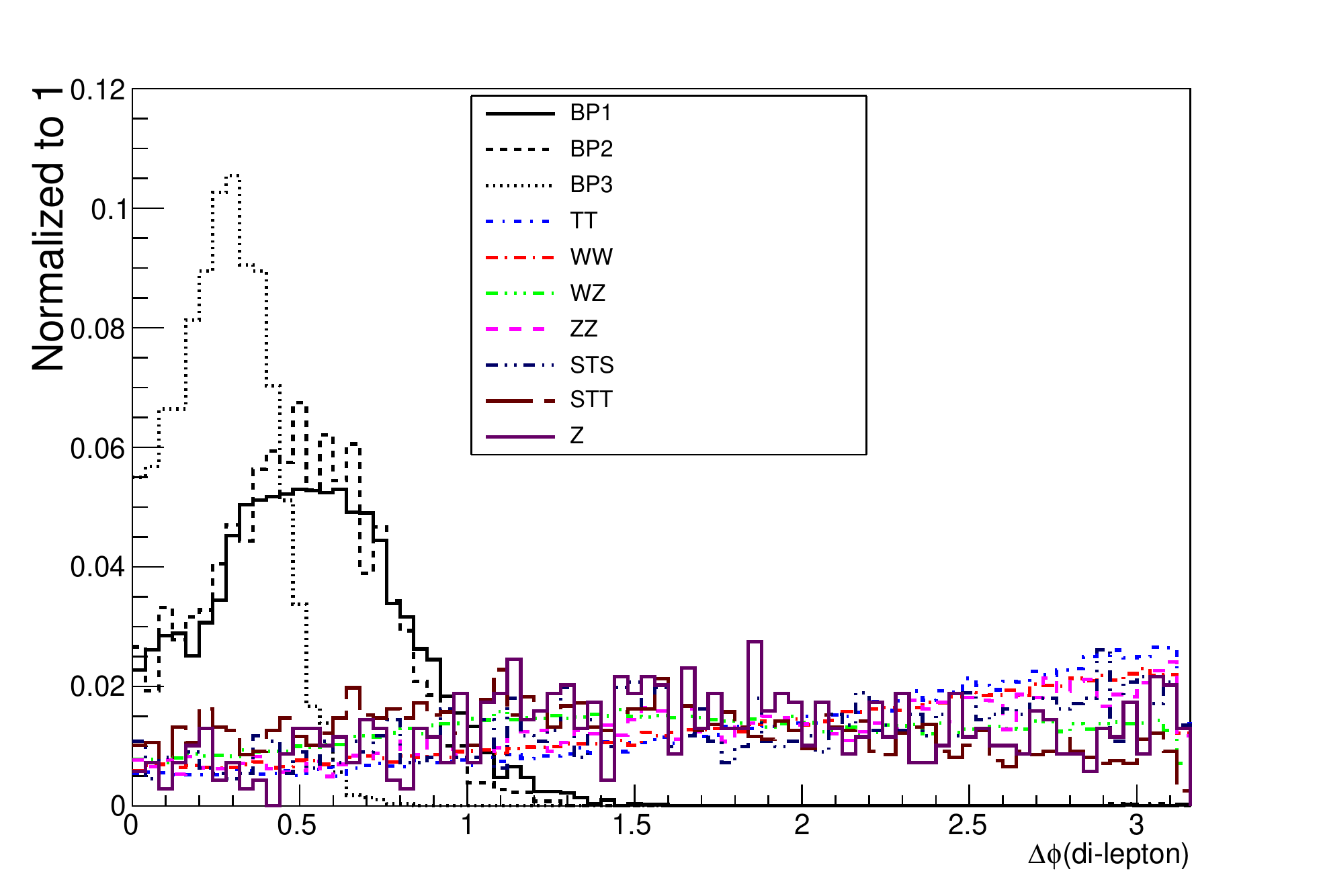}
\\
(a)&(b)
\end{tabular}
\caption{ The distribution of $\Delta\phi$ between the two jets (a)  and two leptons (b) in signal and background events.}
\label{dphi}
  \end{figure*}

The next step is the missing transverse energy (MET) calculation. The MET is calculated by a negative vectorial sum of particles momenta in the transverse plane ignoring neutrinos and scalar $H$ which is set to be stable in {\tt PYTHIA} at event generation time.  Fig. \ref{met2l2j} shows the distribution of MET in signal and background events. As seen in Fig. \ref{met2l2j}, Drell-Yan, $ZZ$ and $WZ$ background events have small MET values while $t\bar{t}$ and $WW$ samples tend to have large MET values due to the leptonic decay of the $W$ boson which produces a pair of lepton plus neutrino. Based on Fig. \ref{met2l2j} a cut on MET is applied as in Eq. \ref{metcut2l2j}. The lower cut is useful for Drell-Yan background suppression and other SM background processes with low MET ($ZZ, WZ, ...$). The upper cut is for $t\bar{t},~WW$ and single top processes which involve $W$ boson leptonic decays resulting in sizable MET in the event: 
\beq
15 ~\textnormal{GeV}~<~\textnormal{MET}~<~50 ~\textnormal{GeV}
\label{metcut2l2j}
\eeq
The azimuthal angle between the jet pair and also the lepton pair is the next quantity to study. As shown on Figs. \ref{dphi}(a) and \ref{dphi}(b), signal events tend to produce collinear di-jet and di-lepton pairs due to the Lorentz boost that $W$ and $Z$ acquire in $H^{\pm} \to W^{\pm}H$ and $A\to ZH$ decays. This is in turn due to the large energy of charged and neutral scalars ($H^{\pm}$ and $A$) in signal events when the hard scattering occurs. Based on Figs. \ref{dphi}(a) and \ref{dphi}(b), a cut is applied on $\Delta\phi$ as in Eq. \ref{dphi2l2j}.
\beq
\Delta\phi(\textnormal{di-jet}) ~<~ 1~~~ \& ~~~\Delta\phi(\textnormal{di-lepton}) ~<~ 1.
\label{dphi2l2j}
\eeq

Before proceeding to the di-lepton invariant mass calculation, we benefit from a feature of signal events which is the fact that there are two pairs in the event. One pair (di-jet) can be used as a clue to reject the Drell-Yan and other SM background processes, and the other pair (di-lepton) can finally be used as the signal signature. 

\begin{figure}
\centering
  \includegraphics[width=0.5\textwidth]{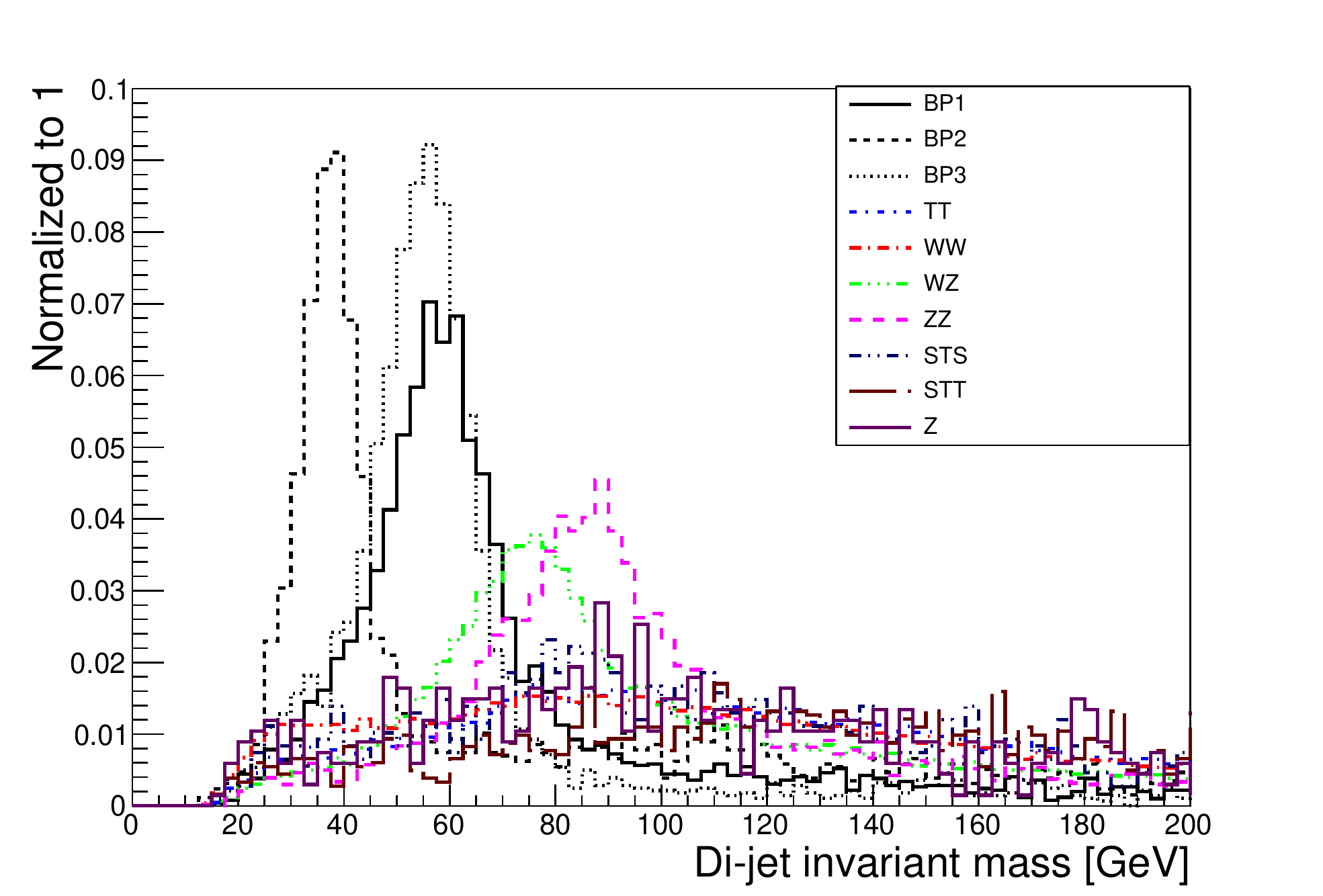}
  \caption{The distribution of di-jet invariant mass in signal and background events.}
  \label{dijet2l2j}
\end{figure}

The di-jet invariant mass distribution at this step shows a reasonable separation between the signal and background processes. Figure \ref{dijet2l2j} shows the distribution of di-jet events in both signal and background events. The single $W$ has been suppressed to a negligible level at this point and is not shown on this plot. The Drell-Yan has to involve a $Z/\gamma\to \ell\ell$ decay to be selected by the di-lepton requirement. Therefore the accompanying jets in the event produce a flat distribution of di-jet invariant mass. The only background events showing peaks near the $W$ or $Z$ are $WZ$ and $ZZ$. Demanding two jets plus two leptons in these events requires $W \to jj$ and $Z\to \ell\ell$ in the first case and $Z_1\to jj$ and $Z_2 \to \ell\ell$ in the second case. Therefore the di-jet invariant mass shows a peak at $\sim$80 GeV in $WZ$ events but a peak at $\sim$90 GeV in $ZZ$ events. A cut on di-jet invariant mass is applied as a ``$W/Z$ veto'' as in Eq. \ref{wveto2l2j}.
\beq
\textnormal{Inv. mass (di-jet)} ~<~60~ \textnormal{GeV}
\label{wveto2l2j}
\eeq

It is now useful to present the selection efficiencies related to all cuts applied up to this point. The selection efficiencies are provided for all signal and background processes for signal significance calculation. Tables \ref{seff} and \ref{beff} show the relative selection efficiencies and the total efficiency of event selection. It should be mentioned that background events have been generated in a final state which is closest to the signal. For example the Drell-Yan efficiency quoted in Tab. \ref{beff} has been obtained from a sample of $Z/\gamma$+jets $\to \ell\ell$+jets. Therefore the number of events at a certain luminosity is obtained by taking the cross section in Tab. \ref{bxsec} multiplied by the BR$(Z \to \ell\ell) \simeq 0.066$, luminosity (300 fb$^{-1}$) and the total efficiency. All Drell-Yan events fall at either very low di-lepton invariant mass values or near the SM $Z$ boson mass at $\simeq$90 GeV. The signal region around 40-50 GeV is thus almost free from the SM background. 

\begin{figure}
\centering
  \includegraphics[width=0.5\textwidth]{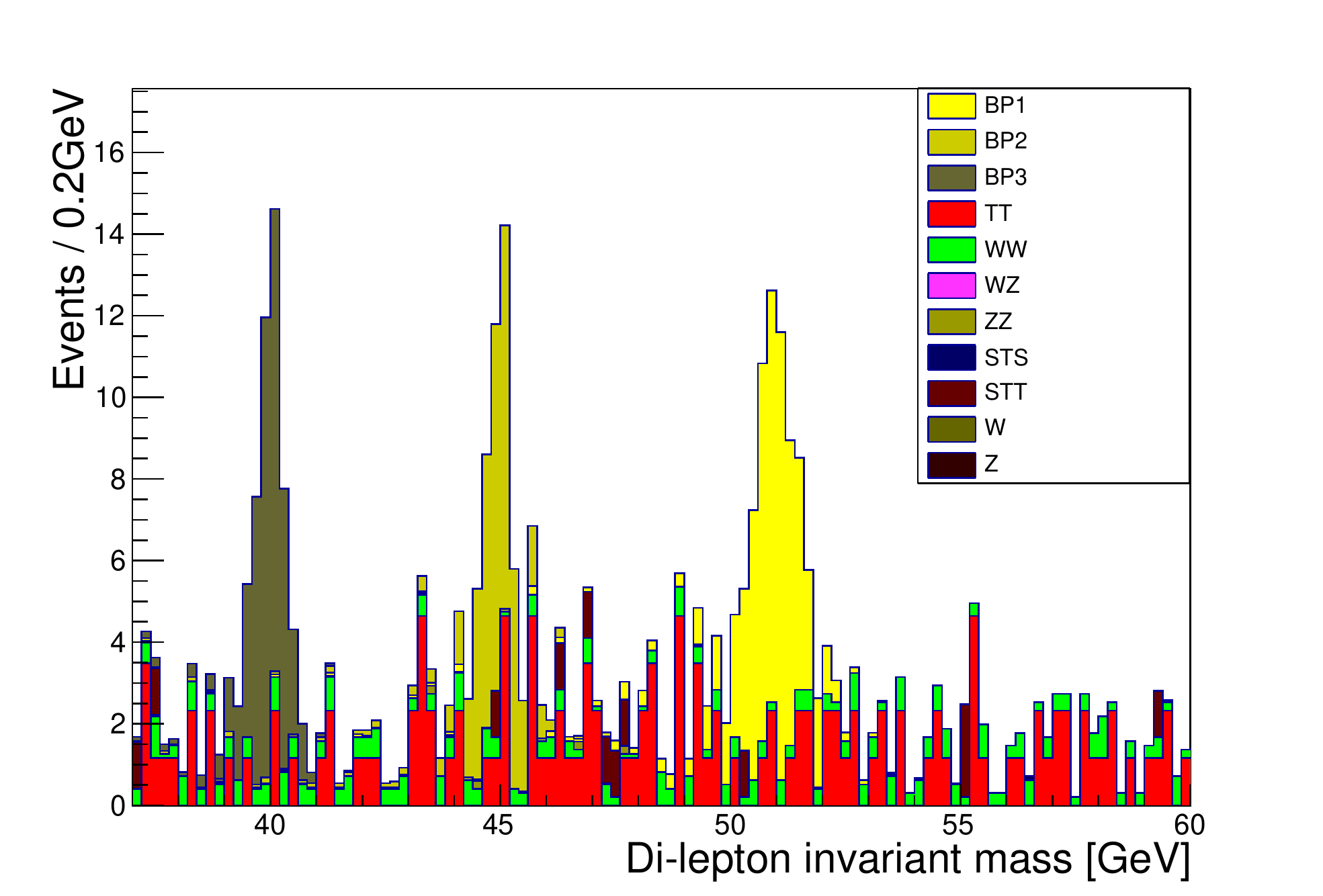}
  \caption{The distribution of di-lepton invariant mass in signal and background events.}
  \label{dilepton2l2j}
\end{figure}

Having applied all cuts (Eq. \ref{eeta2l2j}, \ref{metcut2l2j}, \ref{dphi2l2j} and \ref{wveto2l2j}), the remaining events are used for the di-lepton invariant mass calculation. Figure \ref{dilepton2l2j} shows the distribution of the di-lepton invariant mass in signal and background events. As is seen, signal events are well observable at a high luminosity of 300 fb$^{-1}$ and SM background processes are well under control. 

A mass window on this distribution could determine the signal significance inside the window. On the other hand, one can use a (Gaussian) fit performed on the signal plus background distributions to obtain the off-shell $Z^*$ boson mass which can be used to obtain information about the mass difference between the charged scalar and the $H$ dark matter candidate mass. 

Table \ref{sig2l2j} shows the mass window position and the obtained signal significance in the mass window. Based on these results, a $5\sigma$ discovery is possible in the era of 300 fb$^{-1}$ integrated luminosity. 
\begin{table}
\centering
\begin{tabular}{|c|c|c|c|}
\hline
& \multicolumn{3}{|c|}{$\ell\ell H jj H$}\\ 
Selection cut & BP1 & BP2 & BP3 \\
\hline
2 jets Eq. \ref{eeta2l2j} & 0.27 & 0.17 & 0.27 \\
\hline
2 leptons Eq. \ref{eeta2l2j} & 0.31& 0.27& 0.15\\
\hline
MET Eq. \ref{metcut2l2j}& 0.62& 0.59& 0.70\\
\hline
$\Delta\phi$ Eq. \ref{dphi2l2j}& 0.74& 0.66& 0.77\\
\hline
W/Z veto Eq. \ref{wveto2l2j}& 0.56& 0.82& 0.73\\
\hline
Total eff. & 0.021 & 0.014& 0.016 \\
\hline
\end{tabular}
\caption{Signal selection efficiencies and the total efficiency. \label{seff}}
\end{table}
\begin{table*}[t]
\centering
\begin{tabular}{|c|c|c|c|c|c|c|c|c|}
\hline
& \multicolumn{8}{|c|}{SM Background, $\ell\ell H jj H$ analysis}\\ 
Selection cut & $t\bar{t}$ & WW & WZ &ZZ&STS &STT & W&Z/$\gamma$ \\
\hline
2 jets Eq. \ref{eeta2l2j}& 0.022 &0.046 &0.22 & 0.19 & 0.013 &0.04 & 0.015&0.006\\
\hline
2 leptons Eq. \ref{eeta2l2j}& 0.46&0.4 &0.42 & 0.077 &  0.017& 0.0097 & 0.0003&0.28\\
\hline
MET Eq. \ref{metcut2l2j}&  0.29&  0.39&0.042 &0.037 &0.53 & 0.52 &0.46 &0.042\\
\hline
$\Delta\phi$ Eq. \ref{dphi2l2j}& 0.046&0.077 &0.13 & 0.082 & 0.064& 0.053 & 0&0.10\\
\hline
W/Z veto Eq. \ref{wveto2l2j}& 0.37 &0.42 &0.31 &  0.28 &  0.44&0.39 &0&0.32\\
\hline
Total eff. & 5.02e-05 &2.3e-4 &1.5e-4 & 1.2e-05 &3.1e-06 &4.1e-06 & 0&2.3e-06 \\
\hline
\end{tabular}
\caption{Background selection efficiencies and the total efficiency for each background. STS and STT denote the single top $s$ and $t$ channels respectively. \label{beff}}
\end{table*}
\begin{table*}[t]
\centering
\begin{tabular}{|c|c|c|c|c|c|c|}
\hline
& \multicolumn{6}{|c|}{Signal}\\ 
 & \multicolumn{2}{|c|}{BP1} & \multicolumn{2}{c|}{BP2} & \multicolumn{2}{c|}{BP3} \\
& lower cut & upper cut & lower cut & upper cut & lower cut & upper cut \\
\hline
$m(\ell_1,\ell_2)$ [$\gev$] & 50& 52& 44 & 46 & 39 & 41 \\
\hline
$N_S$ &  \multicolumn{2}{|c|}{64} & \multicolumn{2}{c|}{43} & \multicolumn{2}{c|}{48} \\
\hline
$N_B$&  \multicolumn{2}{|c|}{16} & \multicolumn{2}{c|}{23} & \multicolumn{2}{c|}{12} \\
\hline
$\frac{N_S}{\sqrt{N_B}}$ &  \multicolumn{2}{|c|}{16} & \multicolumn{2}{c|}{9} & \multicolumn{2}{c|}{14} \\
\hline
\end{tabular}
\caption{Mass window cuts in GeV, the number of signal and background events in the mass window and the signal significance at 300 fb$^{-1}$. \label{sig2l2j}}
\end{table*}
\subsection{The trilepton channel}
The trilepton channel is a signal process which involves three leptons in the event. The final goal in the analysis of this channel is to obtain a di-lepton invariant mass distribution like what was obtained in the previous section. There are, however, different combinations of leptons out of the three in each event. 

The analysis starts with a kinematic threshold applied on transverse energies and flight directions of leptons and jets as in Eq. \ref{eeta3l}.
\beq
E_{T}^{\textnormal{jet/lepton}}> 20 ~\textnormal{GeV},~~|\eta|<3.
\label{eeta3l}
\eeq
The signal cross section in this channel is small due to the small branching ratio of leptonic decays of gauge bosons. Therefore the lepton transverse energy has been set to 20 GeV to allow for more signal events to be selected. The jet transverse energy threshold has also been carefully reduced to 20 GeV to increase the number of selected jets. The final aim with the jet analysis is a jet veto. The jet veto may suppress some of signal events due to the initial/final state radiation appeared as jets, however, it suppresses the $t\bar{t}$ and single top backgrounds dramatically. Decreasing the jet energy threshold increases the chance of a jet to be selected. This results in a higher suppression of background events by the jet veto finally. An event is required to contain no jet and exactly three leptons satisfying the requirement in Eq. \ref{eeta3l}.   

\begin{figure}
\centering
  \includegraphics[width=0.5\textwidth]{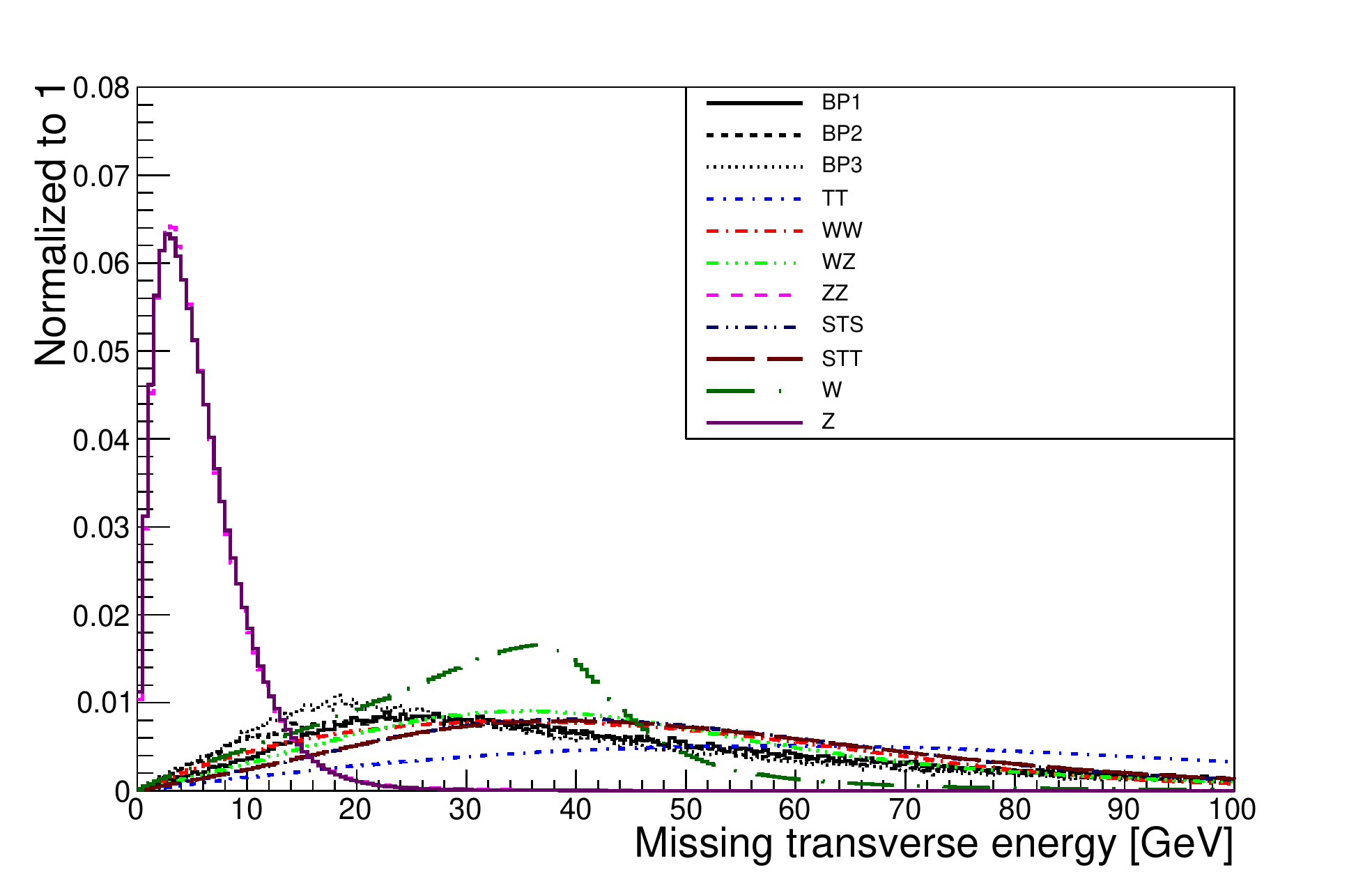}
  \caption{The distribution of missing transverse energy in signal ($\ell\ell H\ell\nu H$) and background events.}
  \label{met3l}
\end{figure}

The missing transverse energy is also calculated with the approach as in the previous section. Figure \ref{met3l} shows the MET distribution based on which a cut is applied as in Eq. \ref{metcut3l}.
\beq
\textnormal{MET} ~>~20 ~\textnormal{GeV}.
\label{metcut3l}
\eeq 

The most powerful approach which can be used to identify the true pair of leptons which come from a $Z$ boson is the analysis of (azimuthal) angles between each pair as well as the angle between each lepton and the neutrino (MET) direction. 

The underlying assumption in this approach is that $A$ and $H^{\pm}$ fly back-to-back after the hard scattering (Fig. \ref{feyn}b). When subsequent decays occur ($A\to ZH$ and $H^{\pm}\to W^{\pm}H$), the $Z$ and $W$ bosons will also be back-to-back and their decay products fly in the same directions as theirs. The flight angle between each lepton from the $Z$ decay and the lepton from the $W$ decay tends to be large (near $\pi$). The lepton and neutrino from the $W$ decay are almost collinear and therefore the same argument applies to the angles between each $Z$ decay product and the missing transverse energy. All above arguments rely on the fact that there is a large Lorentz boost acquired by the $A$ and $H^{\pm}$ scalars in the hard scattering and is in turn transferred to the gauge bosons. Based on the above statements a search among the three leptons is performed to find the best combination satisfying the $\Delta \phi$ requirement as in Eq. \ref{dphicut3l}. The labels are according to the following decay scheme: $Z \to \ell_1\ell_2$ and $W\to \ell_3 \nu$. 

\begin{figure}
\centering
  \includegraphics[width=0.5\textwidth]{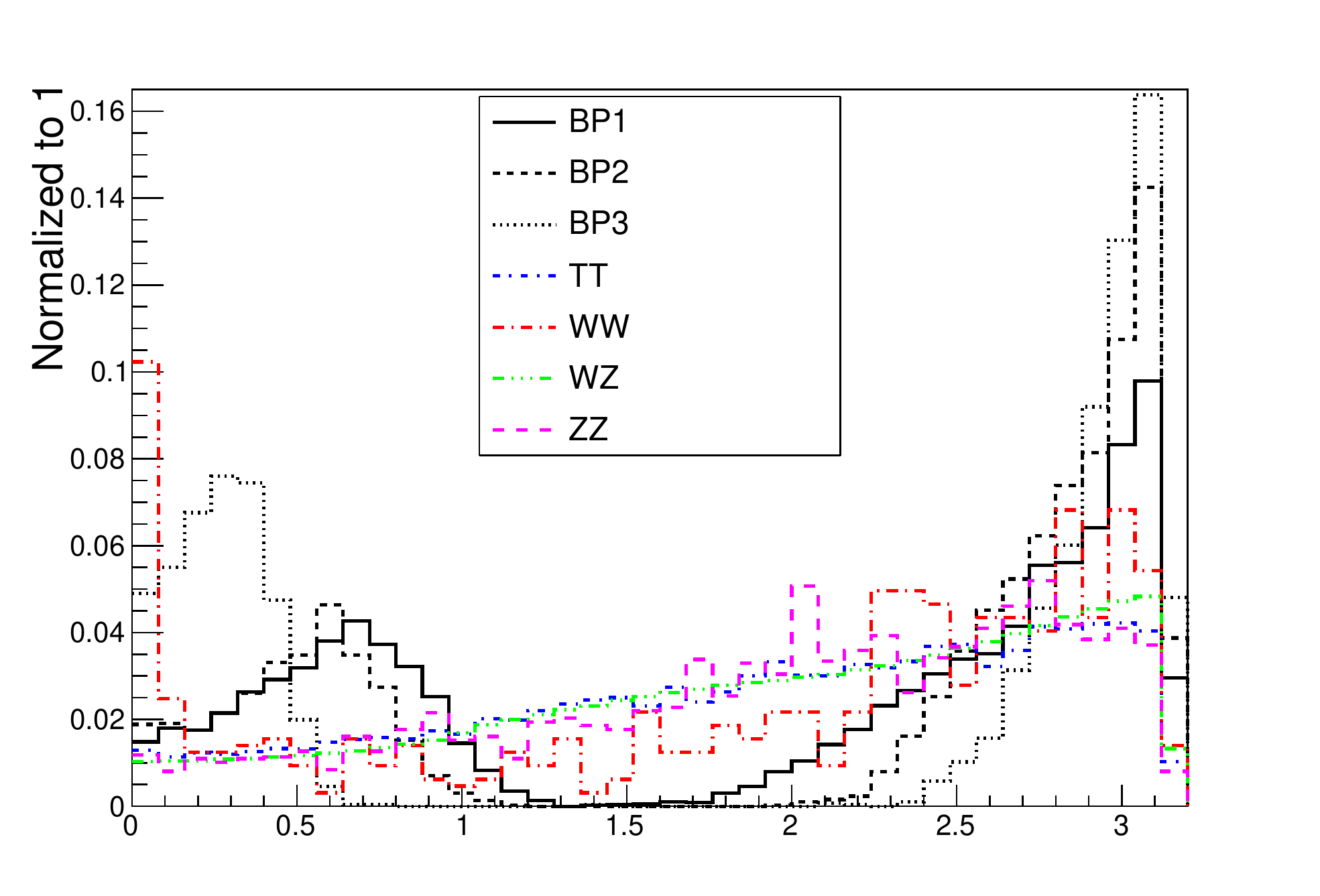}
  \caption{The distribution of $\Delta\phi$ between any lepton pair in signal ($\ell\ell H\ell\nu H$) and background events. Small background samples have been eliminated.}
  \label{dphidilepton3l}
\end{figure}

A random selection of two leptons results in a $\Delta\phi$ distribution as shown in Fig. \ref{dphidilepton3l}. The signal events clearly show two regions of interest. The region with $\Delta\phi<1$ shows the true combination and the region with $\Delta\phi>2$ is related to the wrong combinations. The true or wrong lable here means whether the lepton pair under study are decay products of a $Z$ boson or not. 
\beq
\begin{split}
& \Delta \phi(\ell_1,\ell_3)>2. ~~\&~~ \Delta \phi(\ell_2,\ell_3)>2. \\
& \Delta \phi(\ell_1,MET)>2. ~~\&~~ \Delta \phi(\ell_2,MET)>2. \\
& \Delta \phi(\ell_1,\ell_2)<1.
\end{split}
\label{dphicut3l}
\eeq  
If an event passes the cut on $\Delta\phi$ as in Eq. \ref{dphicut3l}, a ``$W$ veto'' is applied by requiring the transverse mass of the lepton from the $W$ decay ($\ell_3$) and MET to be below the $W$ boson mass as in Eq. \ref{wveto3l}. This cut has been chosen to be hard enough to reject SM background samples such as $WZ$ which are important due to having the same final state as the signal. 
\beq
\textnormal{trans. mass ($\ell_3$,MET)} ~<~60~ \textnormal{GeV}
\label{wveto3l}
\eeq
Now before proceeding to the final distribution which is the di-lepton invariant mass, selection efficiencies have to be calculated for a correct normalization of signal and background samples. Tables \ref{seff2} and \ref{beff2} show relative efficiencies in signal and background events respectively. 

\begin{figure}[t]
\centering
  \includegraphics[width=0.5\textwidth]{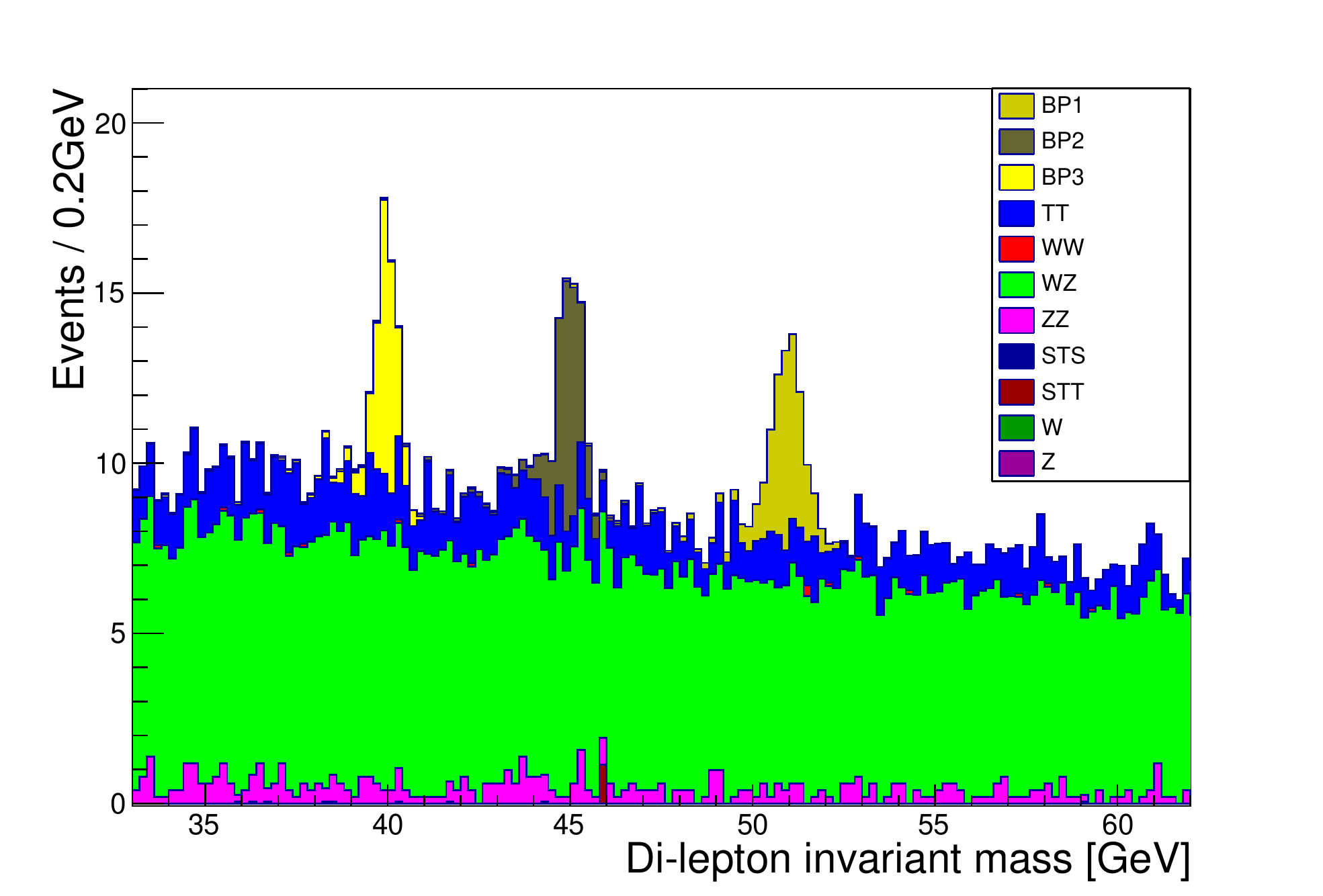}
  \caption{The distribution of di-lepton pair invariant mass in signal ($\ell\ell H \ell \nu H$) and background events.}
  \label{dilepton3l}
\end{figure}

Finally Fig. \ref{dilepton3l} shows the invariant mass of the lepton pair in signal and background events. As is seen, the signal is visible on top of the background events. The gauge boson pair production ($WZ$) is dominated mostly around the $Z$ boson peak at $\simeq$90 GeV and the background distribution in the signal region at 40-50 GeV is almost flat making the signal well visible. The sharp signal peak proves that the $\Delta\phi$ cut has effectively selected the true combination of leptons as the off-shell $Z$ decay products. Table \ref{sig3l} presents the mass window cuts and number of signal and background events at 300 fb$^{-1}$ and the signal significance inside the mass window which corresponds to a 3$\sigma$ evidence. 
\begin{table}
\centering
\begin{tabular}{|c|c|c|c|}
\hline
& \multicolumn{3}{|c|}{$\ell\ell H \ell\nu H$}\\ 
Selection cut & BP1 & BP2 & BP3 \\
\hline
0 jets Eq. \ref{eeta3l} & 0.52 &0.52  & 0.52\\
\hline
3 leptons Eq. \ref{eeta3l} &0.29 &0.20 & 0.17\\
\hline
MET Eq. \ref{metcut3l}&0.76 &0.72 &0.69 \\
\hline
$\Delta\phi$ Eq. \ref{dphicut3l}&0.31 & 0.43 &0.51 \\
\hline
W veto Eq. \ref{wveto3l}&0.90 & 0.97 & 0.96\\
\hline
Total eff. &0.032  & 0.031&0.030  \\
\hline
\end{tabular}
\caption{Signal selection efficiencies and the total efficiency. \label{seff2}}
\end{table}
\begin{table*}[t]
\centering
\begin{tabular}{|c|c|c|c|c|c|c|c|c|}
\hline
& \multicolumn{8}{|c|}{SM Background, $\ell\ell H \ell\nu H$ analysis}\\ 
Selection cut & $t\bar{t}$ & WW & WZ &ZZ&STS &STT & W&Z/$\gamma$ \\
\hline
0 jets Eq. \ref{eeta3l}& 0.011 & 0.59 &0.55 &0.61&0.052 & 0.067&0.75 &0.89  \\
\hline
3 leptons Eq. \ref{eeta3l}& 0.037&0.0001 & 0.42& 0.26& 0.0005& 0.0001 &5.3e-07&1.5e-05\\
\hline
MET Eq. \ref{metcut3l}& 0.91&0.82 & 0.86& 0.005 & 0.87 &0.82&1 &0.07 \\
\hline
$\Delta\phi$ Eq. \ref{dphicut3l}& 0.19 &0.36& 0.23& 0.14&0.27& 0.16&1& 0.3     \\
\hline
W/Z veto Eq. \ref{wveto3l}&0.67  &0.62 & 0.77 & 0.97&0.63 & 0.71&0.5&   1  \\
\hline
Total eff. & 4.8e-05 & 1.5e-05 & 0.035& 0.0001& 3.7e-06 & 1e-06& 2e-07&3e-07 \\
\hline
\end{tabular}
\caption{Background selection efficiencies and the total efficiency for each background. STS and STT denote the single top $s$ and $t$ channels respectively. \label{beff2}}
\end{table*}
\begin{table*}
\centering
\begin{tabular}{|c|c|c|c|c|c|c|}
\hline
& \multicolumn{6}{|c|}{Signal}\\ 
 & \multicolumn{2}{|c|}{BP1} & \multicolumn{2}{c|}{BP2} & \multicolumn{2}{c|}{BP3} \\
& lower cut & upper cut & lower cut & upper cut & lower cut & upper cut \\
\hline
$m(\ell_1,\ell_2)$ [$\gev$] & 50& 52& 44 & 46 & 39 & 41 \\
\hline
$N_S$ &  \multicolumn{2}{|c|}{31} & \multicolumn{2}{c|}{30} & \multicolumn{2}{c|}{28} \\
\hline
$N_B$&  \multicolumn{2}{|c|}{86} & \multicolumn{2}{c|}{99} & \multicolumn{2}{c|}{104} \\
\hline
$\frac{N_S}{\sqrt{N_B}}$ &  \multicolumn{2}{|c|}{3.3} & \multicolumn{2}{c|}{3.0} & \multicolumn{2}{c|}{2.7} \\
\hline
\end{tabular}
\caption{Mass window cuts in GeV, the number of signal and background events in the mass window and the signal significance at 300 fb$^{-1}$. \label{sig3l}}
\end{table*}

\section{Discussion}
\label{Discussion}
Comparing the results of the two analyses, i.e., the di-lepton plus di-jet channel ($\ell\ell HjjH$) and the trilepton channel ($\ell\ell H \ell\nu H$), one may realize that the overall shape of the signal distribution is almost the same. However, there is less background in the di-lepton plus di-jet channel and more signal due to the use of the hadronic decay of the $W$ boson. These reasons result in a higher significance. 

The $WZ$ background plays the more important role in trilepton analysis. However, it has less effect in the di-lepton plus di-jet analysis. The reason is that the signal in di-lepton plus d-ijet channel, has missing transverse energy from the dark matter candidate ($H$). The $WZ$ background should undergo $W\to jj$ and $Z\to \ell\ell$ decays to mimic the signal. There is no real missing transverse energy in such events and this background is well suppressed by the cut on missing transverse energy. 

On the other hand, in the trilepton analysis, the $WZ$ background has to experience $W\to \ell\nu$ and $Z\to \ell\ell$ decays which produce a source of missing transverse energy through the $W$ leptonic decay. This feature of $WZ$ background makes it difficult to suppress in the trilepton analysis resulting in a 3$\sigma$ evidence of the signal in the best case. In di-lepton plus di-jet channel, BP1, BP2 and BP3 will be observable at 5$\sigma$ at integrated luminosity of 300  fb$^{-1}$.  

\section{Conclusions}
\label{Conclusions}
We presented the LHC potential for a dark matter search by introducing an analysis based on inert double model framework. A new set of benchmark points which respect all experimental and theoretical bounds were introduced as working points. The analysis was performed at center of mass energy of 14 TeV. Two main channels, i.e., the di-lepton plus di-jet and trilepton channels were analyzed. Results show that distinguishable signals can be obtained at LHC in the di-lepton plus di-jet channel with significances exceeding $5\sigma$ for 300$\fb^{-1}$.

\acknowledgements
The authors would like to thank Maria Krawczyk  for discussions  and carefully  reading the manuscript, and Aleksander F. Zarnecki for  valuable comments.

\providecommand{\href}[2]{#2}\begingroup\raggedright\endgroup

\end{document}